\documentclass[11pt,twoside]{article}
\usepackage{./asp2014}

\aspSuppressVolSlug
\resetcounters

\begin{document}

\title{Star-forming Filaments and Cores on a Galactic Scale}

\author{James Di Francesco$^1$, Jared Keown$^2$, Rachel Friesen$^3$, Tyler Bourke$^4$,
and Paola Caselli$^5$
\affil{$^1$National Research Council of Canada, Victoria, BC, Canada; \email{james.difrancesco@nrc-cnrc.gc.ca}}
\affil{$^2$University of Victoria, Victoria, BC, Canada; \email{jkeown@uvic.ca}}
\affil{$^3$National Radio Astronomy Observatory, Charlottesville, VA, U.S.A.; \email{rfriesen@nrao.edu}}
\affil{$^4$Square Kilometre Array Organization, Macclesfield, UK; \email{t.bourke@skatelescope.org}}
\affil{$^5$Max Planck Institute for Extraterrestrial Physics, Garching, Germany;\email{caselli@mpe.mpg.de}}
}

\paperauthor{James Di Francesco}{james.difrancesco@nrc-cnrc.gc.ca}{0000-0002-9289-2450}{National Research Council of Canada}{Herzberg Astronomy \& Astrophysics Research Centre}{Victoria}{BC}{V9E 2E7}{Canada}
\paperauthor{Jared Keown}{jkeown@uvic.ca}{0000-0003-2628-0250}{University of Victoria}{Department of Physics and Astronomy}{Victoria}{BC}{V8P 5C2}{Canada}
\paperauthor{Rachel Friesen}{rfriesen@nrao.edu}{0000-0001-7594-8128}{North American ALMA Science Center}{National Radio Astronomy Observatory}{Charlottesville}{VA}{22903-2475}{USA}
\paperauthor{Tyler Bourke}{t.bourke@skatelescope.ca}{}{Square Kilometre Array Organization}{Jodrell Bank Observatory}{Macclesfield}{Cheshire}{SK11 9DL}{UK}
\paperauthor{Paola Caselli}{pcaselli@mpe.mpg.de}{0000-0003-1481-7911}{Max-Planck-Institute for Extraterrestrial Physics}{}{Garching}{}{85748}{Germany}

\begin{abstract}
Continuum observations of molecular clouds have revealed a surprising amount of substructure in 
the form of filaments of a few pc length and cores of $\sim$0.1~pc diameter.  Understanding the 
evolution of these substructures towards star formation requires the kinematic and dynamical insights
provided uniquely by sensitive line observations at high angular and spectral resolution.  In this 
Chapter, we describe how an ngVLA can probe effectively the dynamics of filaments and cores in
nearby star-forming molecular clouds using the NH$_3$ rotation-inversion transitions at 24 GHz.  Such 
emission has been proven to trace well the high column density environments of star-forming cores
and filaments but higher-resolution observations are needed to reveal important details of how dense 
gas is flowing within and onto these substructures.  In particular, we describe how 150 $\times$ 18-m 
antennas with a maximum baseline of 1 km can be used to map sensitively NH$_{3}$ emission across
high column density locations in clouds in roughly an order of magnitude less time than with the current 
Jansky VLA.
\end{abstract}

\section{Introduction}

Stars form out of molecular cloud gas, when dense pockets (i.e., ``cores'') become gravitationally 
unstable and collapse (see Di Francesco et al.\ 2007 for a review).  Recent far-infrared/submillimeter 
continuum observations (e.g., from {\it Herschel} or the JCMT) of Galactic clouds at distances < 3 kpc 
have revealed close connections between the detailed substructures of clouds and star formation.  
First, molecular clouds are suffused with filaments, parsecs-long substructures of $\sim$0.1 pc width, 
regardless of their star-forming activity \citep{Andre10, WT10}.  Second, clouds can have hundreds of 
cores, many of which appear bound and hence likely to form stars in the future \citep{Konyves10}.  
Third, core formation, and hence star formation, appears to be most efficient within supercritical 
filaments above a given column density threshold equivalent to $A_V$ $\sim$ 7 magnitudes 
\citep{Andre10,Andre14}.  Indeed, the resulting core mass function strikingly resembles the initial stellar 
mass function, suggesting the process that forms cores also ultimately determines stellar masses 
\citep{Konyves15}.  Finally, high-mass star and cluster formation occur most efficiently where 
supercritical filaments appear to intersect \citep{Schneider12}.  

Understanding the relationships between filamentary substructures and star formation requires 
kinematic and dynamical insights.  For example, how does gas in clouds assemble into 
filaments and cores?  Also, how will the gas in these substructures evolve?  Is there further coherent
substructure within filaments (e.g., ``fibers''; Hacar et al. 2013, 2017).  The continuum observations 
that identified filaments in molecular clouds do not themselves have the ability to trace their kinematics 
and dynamics.  Instead, observations of line emission are essential to determine how mass flows within 
filaments and cores and whether or not such substructures are stable.  More specifically, to understand 
the structure and dynamical evolution of molecular clouds in general, large-scale surveys of lines must 
be performed at sensitivities and resolutions not possible with currently available observatories.

\begin{figure*}
\begin{center}
\includegraphics[width=0.9\textwidth]{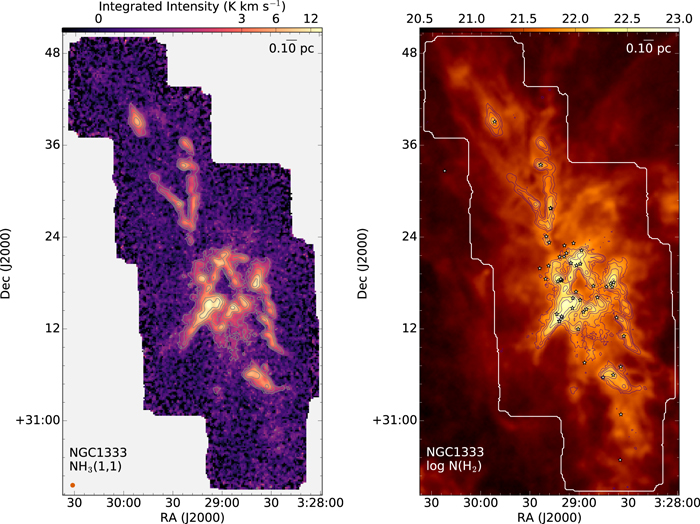}
\end{center}
\caption{\footnotesize
Left: integrated intensity map of the NH$_{3}$ (1, 1) line for the NGC 1333 region obtained by the
Green Bank Ammonia Survey \citep{Friesen17}.  Contours are drawn at [3, 6, 12, 24, ...]-$\sigma$, 
where 1 $\sigma$ $\approx$ 1.0 K km s$^{-1}$ was estimated from emission-free pixels.  Beam size 
and scale bar are shown in the bottom left and top right corners, respectively. Right: the H$_2$ column 
density, log10 [N(H$_2$) cm$^{-2}$], derived from SED fitting of Herschel submillimeter dust continuum 
data (A.\ Singh et al.\ 2018, in preparation).  Contours show the NH$_{3}$ (1,1) integrated intensity, as 
in the left panel.  White contours show the GAS map extent.  Stars show the locations of Class 0/I and
flat-spectrum protostars \citep{Dunham15}.}
\label{fig:fig1}
\end{figure*}

\begin{figure*}
\begin{center}
\includegraphics[width=0.9\textwidth]{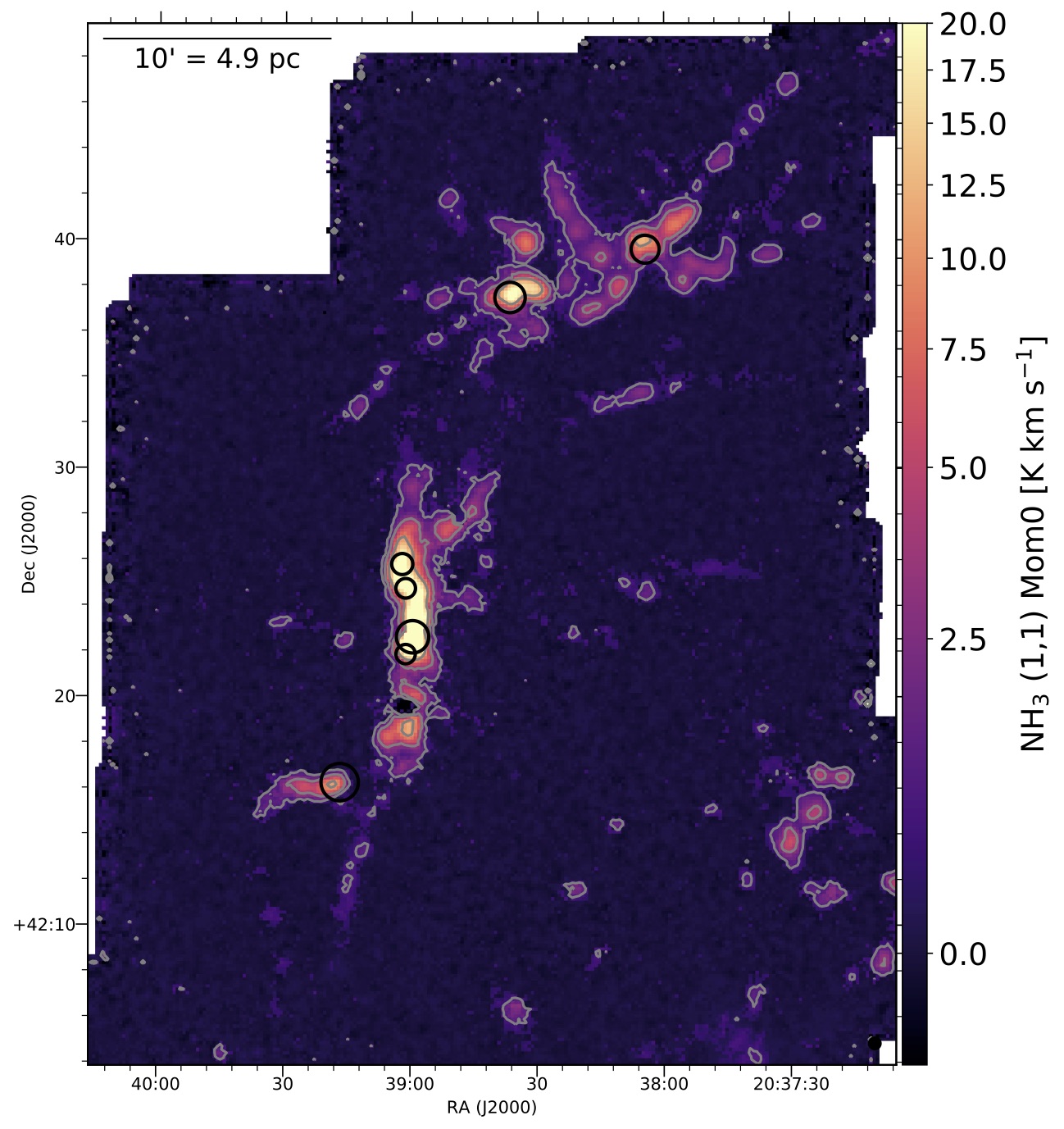}
\end{center}
\caption{\footnotesize
Integrated intensity map of the NH$_{3}$ (1,1) line for the Cygnus X North region, including DR21, 
obtained by the KEYSTONE survey \citep{Keown18}.  Colors range from -1 K km s$^{-1}$ (black) to
10 K km s$^{-1}$ (yellow).  Contours are at 1.0, 3.5, and 10 K km s$^{-1}$.  Circles show the locations 
of H$_{2}$O maser emission simultaneously detected during the NH$_{3}$ observations.  The size of 
the circle indicates the relative brightness of the maser emission.}
\label{fig:fig2}
\end{figure*}

The NH$_3$ rotation-inversion transitions at 24 GHz (e.g., (1,1), (2,2), (3,3), etc.) can probe effectively 
the kinematics and dynamics of star-forming substructures in nearby molecular clouds.  NH$_3$ 
emission has been shown repeatedly to trace best the locations of significant column density revealed 
by continuum emission from dust (e.g., Benson \& Myers 1989).  This behavior follows partly because 
the NH$_3$ transitions are excited in moderately dense gas; e.g., the critical densities of (1,1) and (2,2)
at 10~K are 10$^{3-4}$~cm$^{-3}$ \citep{Ho83}.  In addition, these NH$_3$ transitions have hyperfine 
structure that spreads out their overall optical depths over numerous components, enabling better 
probes of all dense gas along the line of sight \citep{Crapsi07}.  In contrast, other molecules have 
emission that is too optically thick to sample such environments (e.g., $^{12}$CO) or have been 
themselves too drastically depleted in cold dense gas to be effective probes (e.g., C$^{18}$O; see
Di~Francesco et al. 2007).  Even other less-abundant ``dense gas tracers" like HCN or CN emission 
may yet suffer from optical depth and depletion issues on the scales of cores and filaments.  
Furthermore, the hyperfine structure of the NH$_3$ transitions further allows, through simultaneous 
fitting, direct determinations of excitation temperature and  opacity, and hence column density.  Finally, 
these NH$_{3}$ transitions can directly provide the gas kinetic temperature along the line of sight via 
their ratios \citep{Walmsley83}, unlike those of other nitrogen-based molecules (e.g., N$_2$H$^+$).


Given the utility of NH$_3$ emission, it is being widely observed to trace the kinematics and dynamics 
of moderately dense gas in molecular clouds, especially in filaments and cores.  These programs are 
being largely run from the Green Bank Telescope (GBT) using its unique K-band Focal Plane Array 
(KFPA) instrument to map many square degrees of sky in nearby molecular clouds, Giant Molecular 
Clouds, and large swaths of the Galactic Plane.  Figures \ref{fig:fig1} and \ref{fig:fig2} show examples of 
recent wide-field GBT KFPA observations of NH$_{3}$ (1,1) emission toward star-forming molecular 
clouds.  Figure \ref{fig:fig1} (left) shows the integrated intensities of NH$_{3}$ (1,1) emission toward the 
NGC 1333 star-forming region of the Perseus molecular cloud from the Green Bank Ammonia Survey 
with the KFPA \citep{Friesen17}, while Figure \ref{fig:fig1} (right) shows the H$_{2}$ column densities of 
the same region derived from continuum data obtained by {\it Herschel} \citep{Singh18}.  Figure 
\ref{fig:fig2} shows the integrated intensity of NH$_{3}$ (1,1) emission toward the Cygnus X North 
region, including DR21, from the K-band Examinations of Young STellar Object Natal Environments 
(KEYSTONE) survey \citep{Keown18}.  In both cases, widespread NH$_{3}$ emission indicative of 
dense, star-forming gas is seen, much of it in filaments.  

Such wide-field NH$_3$ mapping surveys are already changing our understanding of the stability of 
dense cores and hierarchical structures - and hence, their ability to collapse and form stars, disperse 
without collapse, fragment, and accrete additional mass. For example, \cite{Kirk17} combined
GAS and JCMT Gould Belt Legacy Survey data to estimate the virial states of dense cores within the 
Orion A molecular cloud.  They found that none of the dense cores are sufficiently massive to be bound 
when considering only the balance between self-gravity and the thermal and non-thermal motions 
present in the dense gas, in contrast to analyses which include only thermal gas motions. Instead, most 
of the dense cores are pressure-confined by the additional pressure binding imposed by the weight of 
the ambient molecular cloud material and additional smaller pressure term. Over larger spatial scales, 
\cite{Keown17} showed that NH$_3$-identified structures in Cepheus are gravitationally dominated, 
yet may be in or near a state of virial equilibrium. Filamentary structures in Cepheus have virial 
parameters~$\ll$~2, such that they should be gravitationally unstable unless there is significant support 
by magnetic fields. This result is consistent with a similar analysis of the more massive, more actively 
star-forming Serpens South region \citep{Friesen16}, and hints that large-scale gravitational 
instability is a general property of the dense gas structure in star-forming environments.

Though wide-field observations such as those in Figures \ref{fig:fig1} and \ref{fig:fig2} are enabling 
significant advances to be made about understanding star formation, it remains challenging to recover 
from such data important details of the kinematics and dynamics of the dense gas.  For context, the 
GBT's relatively low angular resolution,i.e., 33$\arcsec$ at 24 GHz, is equivalent to 0.07 pc at the 420 
pc distance of the Orion molecular cloud complex, but cores and filaments each have characteristic 
widths of 0.1 pc \citep{DiFrancesco07,Arzoumanian11}.  Hence, the GBT observations do not resolve 
the gas kinematics and dynamics of such structures in clouds much more distant that Orion.  Indeed,
higher-resolution observations are critical for such targets, to probe for further substructure and allow 
details of mass flow and dynamical stability to be recovered.  Though the Jansky VLA can be currently 
used to observe NH$_{3}$ emission at 24 GHz, its limited point-source sensitivity to line emission, its 
relatively small field-of-view, and its insensitivity to moderate spatial scales, qualities effectively 
hardwired since the beginning of the VLA's operations in 1980, make it challenging to map the internal 
details of multi-pc long filaments and numerous cores.  Instead, a next generation Very Large Array
(ngVLA) with improved sensitivities and wider fields-of-view will enable acquisition of the data needed
to probe best the kinematics and dynamics of dense star-forming gas.  With an flexible ngVLA, targets 
could include large samples of targeted individual pre-/protostellar cores and their host filaments or 
wide-field mapping across star-forming molecular clouds.

\section{NH$_{3}$ observing with ngVLA}

This science is uniquely addressed with sensitive observations of low surface brightness emission
at high spectral and spatial resolution in K-band.  The current ngVLA design will enable acquisition of 
such data through two important features: i) a large amount of surface area over relatively small 
baselines to obtain high sensitivity to compact sources, and ii) a Short Baseline Array (SBA) of 
antennas in a compact fixed configuration that provides high sensitivity to extended sources.  As stated 
above, the Jansky VLA does not have the sensitivities or field-of-view in K-band for wide-field imaging of 
NH$_3$ emission (see also below).  The GBT is useful for locating NH$_3$ emission but it does not 
have the intrinsic angular resolution to resolve cores and filaments beyond Orion, e.g., in several key 
Giant Molecular Clouds like Cygnus X (see Figure \ref{fig:fig2}).   At present, no other single-dish radio 
telescope on the planet is as optimally equipped for wide-field K-band observing than the GBT.   The 
SKA's current specifications on sensitivity only go up to 15 GHz.  While there is a specification for 
aperture efficiency at 20 GHz, the SKA antennas are not being optimized for performance above 15 
GHz.   (We note that design work indicates that performance of the SKA antennas will be good at 25 
GHz but their actual performance remains to be determined.)  ALMA can observe emission from 
N$_2$H$^+$, a nitrogen-based molecule chemically similar to NH$_3$ that can also trace dense gas, 
but the critical density of its lowest rotational transition is 1-2 orders of magnitude higher than those of 
the lower NH$_3$ transitions, suggesting it is less able to map typically lower density filaments or the 
transition regions between dense cores and the filaments within which they are embedded.  Moreover, 
N$_2$H$^+$ cannot be used to determine kinetic gas temperatures directly as NH$_3$ can.   (Note 
that an ngVLA with frequency coverage up to 115 GHz will be able to detect N$_2$H$^+$ 1-0 
emission at 93 GHz at higher sensitivity than ALMA, allowing sensitive probes of compact dense
gas in cores.)  

No other facilities are planned to sample K-band frequencies with the sensitivities and resolutions of
the ngVLA.  Synergies may exist with future far-infrared or submillimeter space missions like SPICA
and OST, if such facilities have wide-field continuum imaging capabilities.  In particular, the potential 
addition of polarizers on these telescopes would provide high-resolution information on magnetic field 
morphologies, from which ngVLA line data could allow magnetic field strengths to be determined using t
the Davis-Chandrasekhar-Fermi method.  Though cores and filaments emit largely at far-infrared and 
submillimeter wavelengths, maps of their internal structures at resolutions higher than {\it Herschel} or 
the JCMT, i.e., 1-3$\arcsec$, could arise from extinction mapping of molecular cloud material using 
deep wide-field near-infrared observations from next-generation facilities such as TMT/E-ELT or LSST.  
Complementarity with kinematic information on such scales is therefore vital.

There are hundreds of known molecular clouds, from those situated relatively near the Sun (0.1-0.5 
kpc distance) to those largely confined to the Galactic Plane ($\sim$10 kpc distance), so there are 
several targets available from any potential ngVLA site.  The targets, however, are generally within 30 
degrees of the Galactic Plane.  Dense gas is found within molecular clouds in relatively isolated 
locations, a small percentage of total cloud surface areas.  Nominally, such locations exhibit extended 
NH$_{3}$ emission over $\sim$10$\arcmin\ \times$ 10$\arcmin$ scales in single-dish maps.  The 
typical peak brightnesses of NH$_3$ (1,1) emission are 1-3 K (T$_{mb}$) in a 33$\arcsec$ beam.  
Assuming no beam dilution, those brightnesses translate to 7.5-25 mJy beam$^{-1}$ for a ``standard 
resolution'' $\sim$4$\arcsec$ beam (see below).   Detection of such emission at an SNR of $\sim$5,
the minimum needed for high-quality line fitting, requires per channel sensitivities of $\sim$5 mJy
beam$^{-1}$.

\begin{figure*}
\begin{center}
\includegraphics[width=0.9\textwidth]{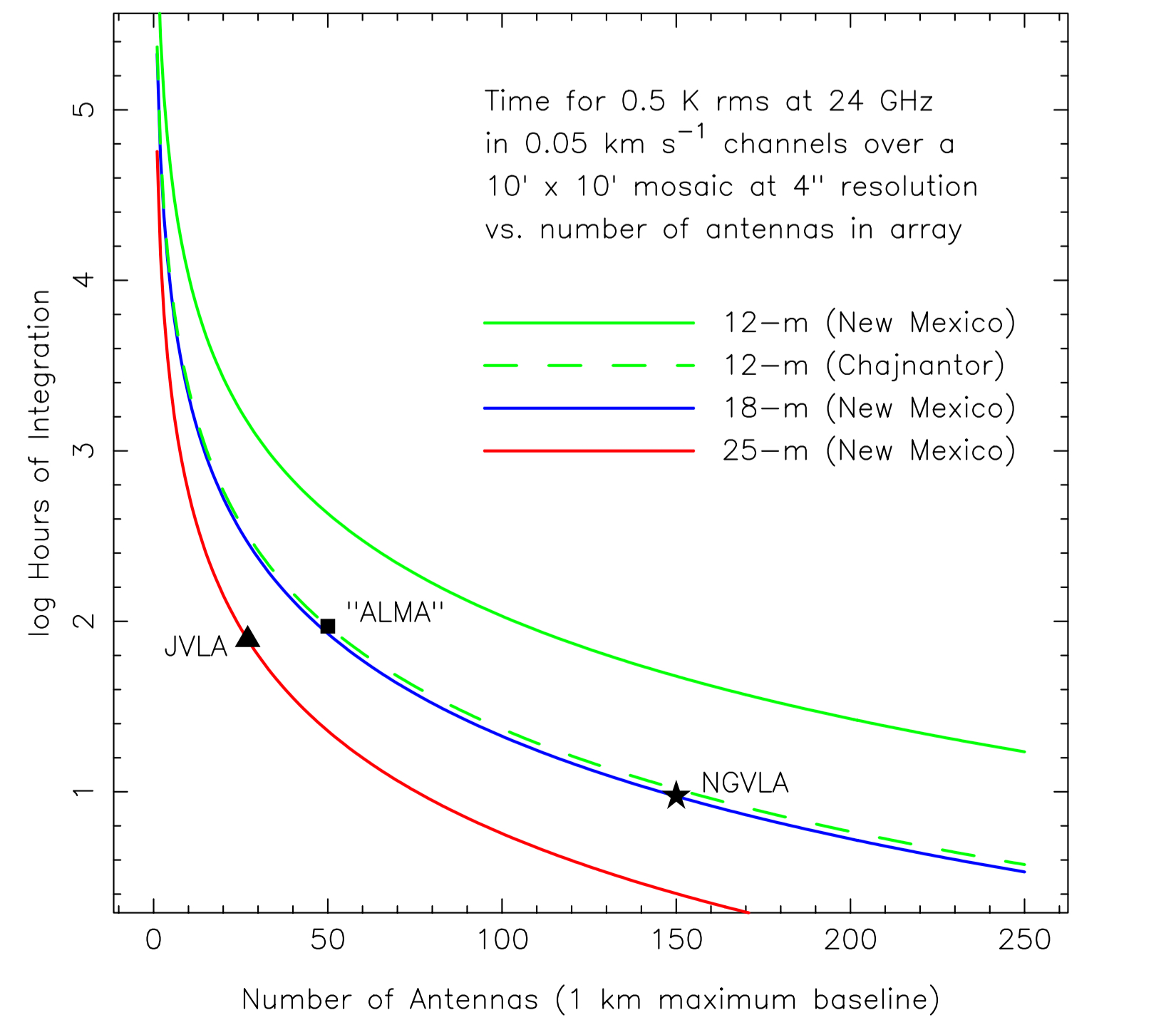}
\end{center}
\caption{\footnotesize
Hours of integration required to reach a 1 $\sigma$ rms = 0.5 K at 24 GHz in a 0.05 km s$^{-1}$ 
channel over a 10$\arcmin$ $\times$ 10$\arcmin$ mosaic at 4$\arcsec$ FWHM resolution given the
number of antennas in an array.  Note that this resolution requirement means the number of antennas 
within about $\sim$1 km maximum baseline.  The number of hours is obtained using the current 
Jansky VLA Exposure Calculator and scaling the integration time by the $D^{2}N^{4}$ scaling
relation for mapping speed.  ($D$ is the antenna diameter and $N$ is the number of antennas.)  The 
solid red line shows the number of 25-m antennas within 1~km needed to reach this target sensitivity 
in New Mexico (i.e., assuming Jansky VLA site conditions).  The black triangle indicates the current 
Jansky VLA of 27 antennas.  The solid green line shows  the number of 12-m diameter antennas within
1~km needed in New Mexico, while the dashed green line shows the number of 12-m ALMA-like 
antennas required at the Llano de Chajnantor in Chile which has superior site conditions.  For 
reference, the black square indicates the 50 $\times$ 12-m antennas of ALMA, though ALMA has no
plans for K-band receivers for the foreseeable future.  The solid blue line shows the number of 18-m 
antennas within 1~km needed to reach the target sensitivity in New Mexico, with a star indicating the 
reasonable number (150) needed in a compact distribution of ngVLA antennas.}
\label{fig:fig3}
\end{figure*}

\section{Example ngVLA Mosaic Observations}

To determine examples of reasonable observations for an ngVLA, we first assume a standard target
resolution of 4$\arcsec$ FWHM, similar to that provided by the D configuration of the Jansky VLA.  
Going to higher resolution has a profound impact on surface brightness sensitivity, so there are 
tradeoffs expected between sensitivity and field-of-view.  Of course, more distant clouds for which 
higher resolution may be most beneficial will be intrinsically more compact and require fewer pointings 
per mosaic.  Single-dish data show NH$_3$ emission in dense gas regions of nearby molecular clouds 
to be extended over $\sim$10$\arcmin$ $\times$ 10$\arcmin$ fields, leading to a mapped image size
of $\sim$100 sq. arcmin.  In addition, the target maximum angular scale is $\sim$30$\arcsec$.  The 
important addition of an SBA to the ngVLA will enable low surface brightness emission to be detected 
over larger scales.  Here, however, we focus on ngVLA observations without an SBA, to provide a more 
straightforward comparison with the current Jansky VLA.  

We expect that a target sensitivity of $\sim$0.5 K in channels 0.05 km s$^{-1}$ wide over a mosaic of 
10$\arcmin$ $\times$ 10$\arcmin$ provides a reasonable minimum goal for ngVLA observations of 
dense gas in star-forming regions.  Such high spectral resolution is necessary to resolve NH$_{3}$ 
lines at low temperatures, e.g., the thermal line width of NH$_{3}$ at 10 K is 0.07 km s$^{-1}$.  
According to the VLA Exposure Calculator\footnote{https://obs.vla.nrao.edu/ect/}, 24 GHz can be 
observed in a single pointing to 0.5 K sensitivity in 0.05 km s$^{-1}$  channels with 25 antennas in D 
configuration (i.e., $\sim$3-4$\arcsec$ FWHM resolution) in $\sim$1.2 hours of winter time on source
or $\sim$2.3 hours with overheads.  (Baselines of up to 4 km are needed to reach 1$\arcsec$ 
resolution.)  A $\sim$10$ \arcmin$ $\times$ 10$\arcmin$ field of dense gas, however, requires $\sim
$126 pointings for Nyquist sampling in a mosaic.  Indeed, a mosaic may benefit from a $\sqrt{2}$ 
increase in sensitivity from overlapping pointings, equivalent to a savings of time by a factor of 2.  
Accordingly, the Jansky VLA can reach the target sensitivity over a 10$\arcmin$ $\times$ 10$\arcmin$ 
mosaic in approximately 78 hours.  Though tractable for a small cloud of 100 sq. arcmin.\ size or
less, mapping the NH$_3$ emission of larger clouds with the Jansky VLA would take numerous such 
mosaics to cover their entirety, and hence would require a prohibitively large amount of observing time.  
For example, the mapped region of Cygnus X shown in Figure \ref{fig:fig2} consists of eleven such 
10$\arcmin$ $\times$ 10$\arcmin$ fields.

Figure \ref{fig:fig3} shows the number of hours needed to reach the target sensitivity of 0.5~K in
0.05 km~s$^{-1}$ channels over a 10$\arcmin$ $\times$ 10$\arcmin$ mosaic for various numbers of
antennas of different size.  The trends follow the well-known $D^{2}N^{4}$ scaling relation for mapping 
speed, where $D$ and $N$ are the diameter and number of the antennas, respectively.  The hours 
needed for the Jansky VLA of 27 antennas of 25-m diameter is indicated, as is the impact on observing 
time given a lesser or greater number of such antennas.  Also shown in Figure \ref{fig:fig3} are curves 
indicating the observing times needed to reach the target sensitivity for 18-m and 12-m diameter
antennas in New Mexico, again following the $D^{2}N^{4}$ scaling relation.  For further comparison,
a curve is also shown in Figure \ref{fig:fig3} for ALMA-like 12-m antennas on the Llano de Chajnantor
in Chile, which has site conditions superior to those in New Mexico (e.g., a 50\% better atmospheric 
opacity at 24 GHz is assumed).  ALMA has 50 $\times$ 12-m antennas and the specific hours of 
integration needed for it to reach the target sensitivity is indicated in Figure \ref{fig:fig3}.  Note,
however, that there are no plans to install K-band receivers on ALMA for the foreseeable future.
Also, the sensitivity performance of ALMA equipped with K-band receivers would be merely
comparable to that of the Jansky VLA.  

Figure \ref{fig:fig3} indicates that $\sim$150 $\times$ 18-m antennas within 1~km 
distance would be sufficient to improve the observing speed of observing such a mosaic to the target 
sensitivity by roughly an order of magnitude, i.e., 9.4 hours.  In contrast, hundreds of 12-m antennas in 
New Mexico would be needed to reach a similar performance improvement.   Note that for the ngVLA 
curves in Figure \ref{fig:fig3} we assumed current Jansky VLA receiver and antenna performances
and did not include further sensitivity improvements from receiver upgrades or better dish surfaces.

We conclude that 150 antennas with a maximum baseline of $\sim$1 km are required to fulfill
this science case.  Such antenna numbers are needed as an absolute minimum to enable roughly an 
order of magnitude improvement over the current Jansky VLA.  Compact placement of ngVLA
antennas is critical for the ability to image low surface brightness emission from dense gas.  
For context, a circular area of 1-km diameter has the same surface area as $\sim$3100 antennas of
18-m diameter, so the required configuration will be tight but doable.   At present, however, the ngVLA 
concept includes $\sim$220 $\times$ 18-m antennas of which only $\sim$40\%  (i.e., 88) would be 
located within 2~km and another $\sim$40\% located within 50~km.  Hence, it may be worthwhile for 
the inner ngVLA antennas to have some limited reconfigurability for low surface brightness imaging.  
Note that the inclusion of an SBA in the ngVLA concept will provide complementary sensitivity to low 
surface brightness emission on large scales.  Indeed, a subset of a few SBA antennas equipped with
K-band focal plane arrays to obtain total power observations efficiently would be very beneficial for
this science. Detailed simulations of ngVLA observations of dense gas structures, including inputs from 
various SBA designs, are urgently needed.

\acknowledgements JDF and JK acknowledge the support of a Discovery Grant by the 
Natural Science and Engineering Council of Canada. 


\begin{thebibliography}{}

\bibitem[Andr\'e et al.\ (2014)]{Andre14} Andr\'e, Ph., Di Francesco, J., Ward-Thompson, D., et al. 2014, in Protostars and Planets VI, ed. H. Beuther, R. S. Klessen, C. P. Dullemond, \& T. Henning, (University of Arizona Press, Tucson), p. 27-51
\bibitem[Andr\'e et al.\ (2010)]{Andre10} Andr\'e, Ph., Men'shchikov, A., Bontemps, S., et al.\ 2010, \aap, 518, L102
\bibitem[Arzoumanian et al.\ (2011)]{Arzoumanian11} Arzoumanian, D., Andr\'e, Ph., Didelon, P., et al.\ 2011, \aap, 529, L6
\bibitem[Benson \& Myers (1989)]{Benson89} Benson, P., \& Myers, P. C. 1989, \apjs, 71, 89
\bibitem[Crapsi et al.\ (2007)]{Crapsi07} Crapsi, A., Caselli, P., Walmsley, M., et al. 2007, \aap, 470, 221
\bibitem[Di Francesco et al.\ (2007)]{DiFrancesco07} Di Francesco, J., Evans, N. J., II., Caselli, P., et al.\ in Protostars and Planets V, ed. B. Reipurth, D. Jewitt \& K. Keil, (University of Arizona Press, Tucson), p. 17
\bibitem[Dunham et al.\ (2015)]{Dunham15} Dunham, M. M., Allen, L. E., Evans, N. J., II, et al.\ 2015, \apjs, 220, 11
\bibitem[Friesen et al.\ (2017)]{Friesen17} Friesen, R. K., Pineda, J. E., Rosolowsky, E., et al.\ 2017, \apj, 843, 63
\bibitem[Friesen et al.\ (2016)]{Friesen16} Friesen, R. K., Bourke, T. L., Di Francesco, J., et al.\ 2016, \apj, 833, 204
\bibitem[Hacar et al.\ (2017)]{Hacar17} Hacar, A., Tafalla, M., \& Alves, J.\ 2017, \aap, 606, 24
\bibitem[Hacar et al.\ (2013)]{Hacar13} Hacar, A., Tafalla, M., Kauffmann, J., \& Kov\'acs, A.\ 2013, \aap, 554, 55 
\bibitem[Ho \& Townes (1983)]{Ho83} Ho, P. T. P., \& Townes, C. H. 1983, \araa, 21, 239
\bibitem[Keown et al.\ (2018)]{Keown18} Keown, J., Di Francesco, J., Rosolowsky, E., et al.\ 2018, in preparation
\bibitem[Keown et al.\ (2017)]{Keown17} Keown, J., Di Francesco, J., Kirk, H., et al.\ 2017, \apj, 850, 3
\bibitem[Kirk et al.\ (2017)]{Kirk17} Kirk, H., Friesen, R. K., Pineda, J. E., et al.\ 2017, \apj, 846, 144
\bibitem[K\"onyves et al.\ (2015)]{Konyves15} K\"onyves, V., Andr\'e, Ph., Men'shchikov, A. et al.\ 2015, \aap, 584, A91
\bibitem[K\"onyves et al.\ (2010)]{Konyves10} K\"onyves, V., Andr\'e, Ph., Men'shchikov, A., et al.\ 2010, \aap, 518, L106
\bibitem[Schneider et al.\ (2012)]{Schneider12} Schneider, N., Csengeri, T., Hennemann, M., et al.\ 2012, \aap, 540, L11
\bibitem[Singh et al.\ (2018)]{Singh18} Singh, A., et al.\ 2018, in preparation
\bibitem[Walmsley \& Ungerechts (1983)]{Walmsley83} Walmsley, M., \& Ungerechts, H.\ 1983, \aap, 122, 164
\bibitem[Ward-Thompson et al.\ (2010)]{WT10} Ward-Thompson, D., Kirk, J., Andr\'e, Ph., et al.\ 2010, \aap, 518, L92

\end{thebibliography}


\end{document}